\documentclass[preprint]{aastex62}

\received{March 7, 2018}
\revised{April 23, 2018}
\accepted{April 24, 2018}
\submitjournal{ApJ}

\shorttitle{HD 141569}
\shortauthors{White \& Boley}

\begin{document}

\title{Extended Millimeter Emission in the HD 141569 Circumstellar Disk Detected with ALMA}

\correspondingauthor{Jacob Aaron White}
\email{jawhite@phas.ubc.ca}

\author[0000-0002-0786-7307]{Jacob Aaron White}
\author{A. C. Boley}
\affil{Department of Physics \& Astronomy \\
University of British Columbia \\
6224 Agricultural Rd. \\
Vancouver, BC V6T 1Z1, CAN}

\begin{abstract}

We present archival ALMA observations of the HD 141569 circumstellar disk at 345, 230, and 100 GHz.  These data detect extended millimeter emission that is exterior to the inner disk.  We find through simultaneous visibility modeling of all three data sets that the system's morphology is described well by a two-component disk model. The inner disk ranges from approximately 16 to 45 au with a spectral index of 1.81 ($q=2.95$) and the outer disk ranges from 95 to 300 au with a spectral index of 2.28 ($q=3.21$). Azimuthally averaged radial emission profiles derived from the continuum images at each frequency show potential emission that is consistent with the visibility modeling. The analysis presented here shows that at $\sim5$ Myr HD 141569's grain size distribution is steeper, and therefore more evolved, in the outer disk than in the inner disk.

\end{abstract}


\section{Introduction}

HD 141569 is a widely studied system that hosts an intricate circumstellar disk of gas and dust centered around a Herbig Ae/Be star. As seen in scattered light images, the dust exhibits large spiral features extending from $\sim 100-400$ au \citep{ancker,weinberger, clampin}. In addition, there is a separate disk interior to the spiral structure. The inner disk was observed in scattered light with Keck/NIRC2 \citep{currie}, which found a disk radius of $\sim39$ au, and $870~\mu$m continuum with ALMA \citep{white16}, which found a radius of $\sim 55$ au. SED modeling further predicts a central clearing in the inner disk around 15 au \citep{maaskant}, although this has yet to be directly observed at any wavelength. 

A small amount of gas was detected between 10 and 50 au through IR observations \citep{brittain, goto}, and ALMA imaged significant CO(3-2) extending from $\sim25$ au to over 200 au, with a potential East-West asymmetry \citep{white16}. The total gas mass has been inferred to be between 2 - 200 M$_{\oplus}$, based on CO observations  \citep{zuckerman95, thi, flaherty, white16}. These estimates are largely dependent on modeling assumptions and adopt an interstellar CO/H$_{2}$ abundance of $\sim10^{-4}$ (which may not be an appropriate ratio for this disk).

The current evolutionary stage of HD141569's disk is unclear. \cite{clampin} describe the spiral features seen with \textit{HST} in the outer disk as a debris disk, but the extent and radial profile is consistent with a protoplanetary disk \citep{flaherty}. Scattered light observations \citep[e.g.,][]{brittain, currie} claim that the inner disk is consistent with a residual protoplanetary disk, but ALMA observations \citep{white16} of the $\sim1$ mm grains and CO find that the gas and dust abundances could be consistent with collisional production. The system’s high CO flux density relative to its continuum may suggest that the disk has significant primordial gas \citep{pericaud}.  However, the short photodissociation timescale for the CO gas suggests that the gas has a non-negligible second-generation component \citep{white16}. As of yet, no planets have been detected in the system, but they could contribute to the observed morphology in the outer disk \citep{augereau}. Disk-disk interactions, such as the photoelectric instability \citep{richert}, can also give rise to spiral structure in opticially thin gas disks such as HD 141569. If the disk is in a debris-like state, then dynamical interactions should also produce mm grains in both the inner and outer disks. The grain size distribution, q, as traced by the spectral index, is expected to be steeper for a debris disk than for a protoplanetary or transitional disk \citep{wyatt, macgregor}, although some overlap is possible. VLA 33 GHz observations sought to constrain the dynamical state of HD 141569's disk, but were potentially biased by unconstrained stellar emission \citep{macgregor, white17}.

In this paper, we present evidence for the presence of mm grains exterior to the inner disk in the HD 141569 system. In Section 2, we describe the data and observations.  The model fitting and their implications are discussed in Sections 3 and 4, respectively.  The results are summarized in section 5.

\section{Observational Data}

Our analysis uses three archival ALMA data sets and a PSF subtracted HST scattered light image from \cite{konishi}. The ALMA observations are at 345 GHz (ID 2012.1.00698.S), 230 GHz (ID 2015.1.01600.S), and 100 GHz (ID 2013.1.00883). The 345 and 100 GHz data were published previously by \citet{white16} and \citet{white17}, respectively. The 230 GHz ALMA data were retrieved from the ALMA archive.

The ALMA 230 GHz observations were taken on 2016 May 16$^{\rm th}$. The total integration time was 20.5 minutes, with about 2.25 minutes on source.  Four different spectral windows (SPW) were used. One SPW had a bandpass of 468.75 MHz centered at 219.565 GHz, one SPW had a bandpass of 117.18 MHz centered at 220.40 GHz, and two SPWs had 2000 MHz bandpasses centered at rest frequencies of 231.02 and 233.02 GHz.  The data were reduced using the Common Astronomy Software Applications ({\scriptsize CASA 4.5.3}) pipeline \citep{casa}, which included WVR calibration; system temperature corrections; bandpass and phase calibration with quasars J1517-2422 and J1549+0237, respectively; and flux calibration with Titan.
 
The 230 GHz observations achieve a sensitivity of $130~\mu \rm Jy~beam^{-1}$. The size of the resulting synthesized beam  is $1.03 \times 0.99$ arcsec$^{2}$ at a position angle of 70.8$^{\circ}$, corresponding to $\sim 112$ au at the system distance of 111 pc.

\section{Visibility Fitting}

Visual inspection of the continuum images of HD 141569 at 345 and 100 GHz did not show any obvious structure outside of the inner disk \citep{white16, white17}. However, continuum observations at $\sim350$ GHz have found a range of flux densities for the system including $3.8\pm0.5$ mJy with ALMA \citep{white16}, $8.2\pm2.4$ mJy with SMA \citep{flaherty}, and $12.6\pm4.6$ mJy with APEX \cite{nilsson}. These observations all have significantly different beam sizes (42, 156, and 2000 au, respectively), and the diffuse flux may not have been fully recovered in previous analyses.

ALMA should in principle be sensitive to mm emission beyond the inner disk, if present. Such emission however may not be apparent in the images, requiring analysis of the visibilities. While previous visibility modeling used a single component model with a uniformly illuminated disk \citep{white16, white17}, our analysis here fits single and two-component power law models to all three ALMA visibility data sets simultaneously.  

The visibility data for each frequency is annularly averaged in 10-k$\lambda$ bins (see Fig.\,\ref{vis1}). The uncertainties represent the standard deviation of each bin. The 230 GHz data have the largest uncertainties due to the relatively short on-source integration time. We adopt a Metropolis-Hastings MCMC modeling approach to explore parameter space using the framework laid out in \citet{white_fom}. For a given set of parameters, we calculate the visibilities and project them to the disk geometry \citep[$\rm PA=356.6^{\circ}$ and $\rm inc=53.4^{\circ}$,][]{white16}. Each model is compared to the data and a $\chi^{2}$ is calculated for each frequency. The three are averaged together with equal weighting to get a representative $\chi^{2}$. We note that since each ALMA data set has a different beam size and sensitivity, an equal weighting of each frequency may not directly reflect the contribution from each set of visibilities.

\subsection{Single-Component Model}

The single-component model assumes that the system is best characterized by a single disk. Previous modeling of the disk at 345 and 100 GHz used the \textit{uvmodelfit} task in {\scriptsize CASA} to fit a uniform surface brightness disk to the visibilities. If the disk has a varying surface brightness, does not extend all the way to the star, or has multiple components, then this previous simple model could underpredict the total flux. We confirm that our MCMC modeling approach, when adopting a uniform surface brightness disk model, recovers results that are consistent with those of \textit{uvmodelfit}.

\begin{figure}
\centering
\includegraphics[width=\textwidth]{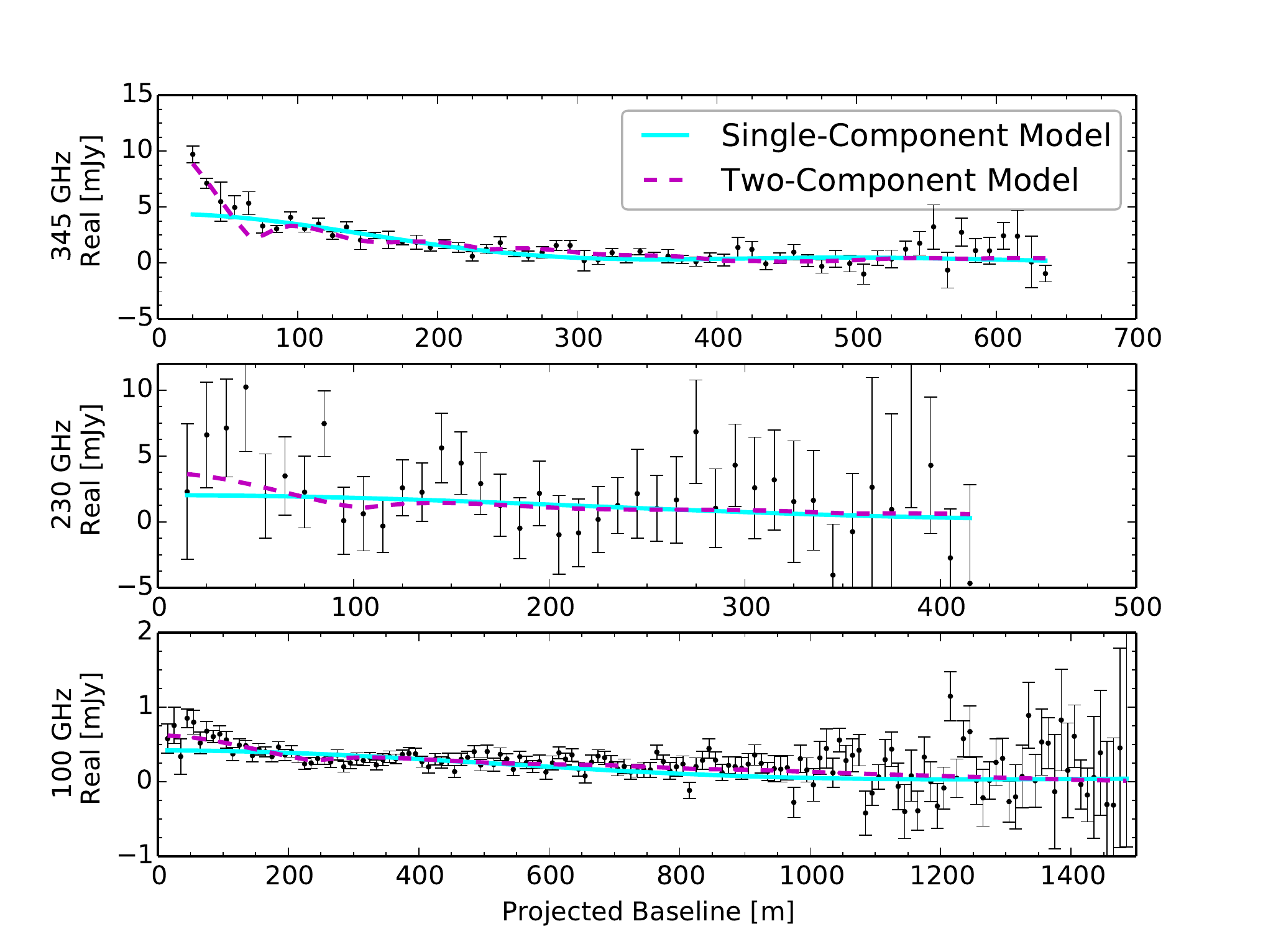}
\caption{Visibility plots of the three ALMA data sets. The real component is shown as a function of projected baseline. The cyan curves are the best fit single-component models for each frequency and the magenta curves are the best fit two-component models for each frequency. The visibility data were annularly averaged with 10-k$\lambda$ bins and the uncertainties shown are the standard deviation of each bin.
\label{vis1}}
\end{figure}

\begin{figure}
\centering
\includegraphics[width=\textwidth]{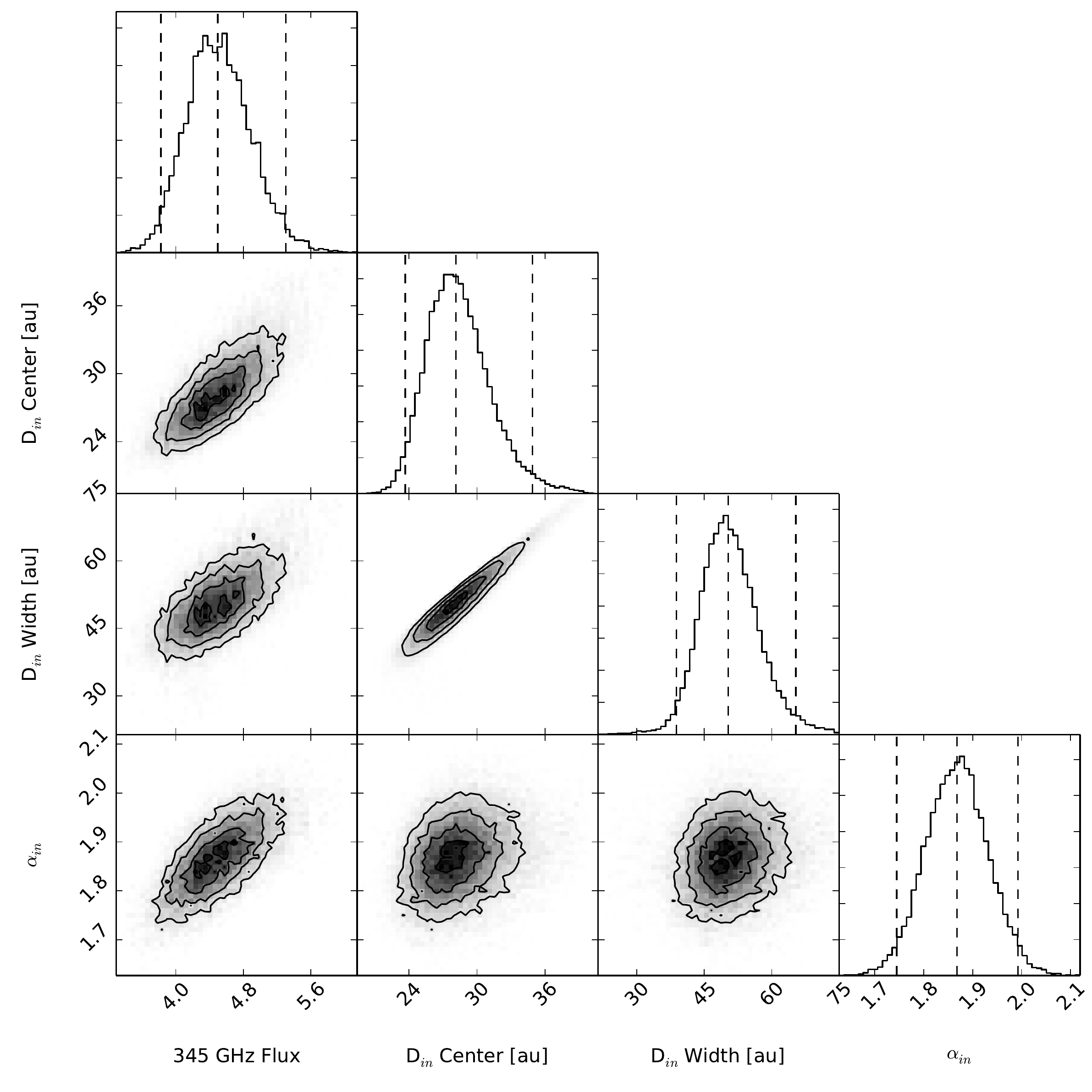}
\caption{MCMC posterior distributions for the single-component model. The dashed lines indicate the most probable values and the 95\% Credible Region. The apparent correlation between the disk center and width is due to the model fitting strongly preferring a disk inner edge near $\sim5$ au.
\label{pdf1}}
\end{figure}

The single-component model used here adopts a surface brightness profile $\propto r^{-1}$ and assumes the disk has an inner and outer edge. The 345 GHz flux, spectral index, disk center, and disk width are free parameters. By fitting the spectral index, the flux at 230 and 100 GHz can be obtained without fitting them directly (and assuming the flux is well defined by a single power law in frequency), reducing the number of free parameters in the modeling process. A uniform prior distribution is chosen for the flux to range from .01 to 50 mJy, the disk from 0.1 to 300 au, and the spectral index from 0 to 4. While the uniform priors span orders of magnitude, and are therefore biased against small parameter values, the flux is expected to be greater than 1 mJy and the inner disk edge is expected to be greater than 1 au (from SED modeling).  Therefore, using very low values for the lower bounds of the flux and the disk  are not expected to affect the results.  To check this, we ran MCMC modeling for the single component using Jeffreys priors, which resulted in negligible differences. The best fit model is given by the cyan curves in Fig.\,\ref{vis1} and the posteriors are shown in Fig.\,\ref{pdf1}. A summary of the most probable model parameters, 95\% Credible Region, and reduced $\chi^{2}$ at each frequency are given in Table \ref{obs}.

The resulting best fit disk extends from about 3 to 53 au with a spectral index of 1.87. The outer edge of the disk is consistent with previous ALMA models of the inner disk and the spectral index is consistent within the uncertainties of the value reported in \citet{white17}. The inner edge of the disk (which is not resolved in the images) is inconsistent with the SED predicted central clearing. The best fit 345 GHz flux is 4.5 mJy and the spectral index gives fluxes of 2.1 and 0.45 mJy at 230 and 100 GHz, respectively.

\subsection{Two-Component Model}

\begin{figure}
\centering
\includegraphics[width=\textwidth]{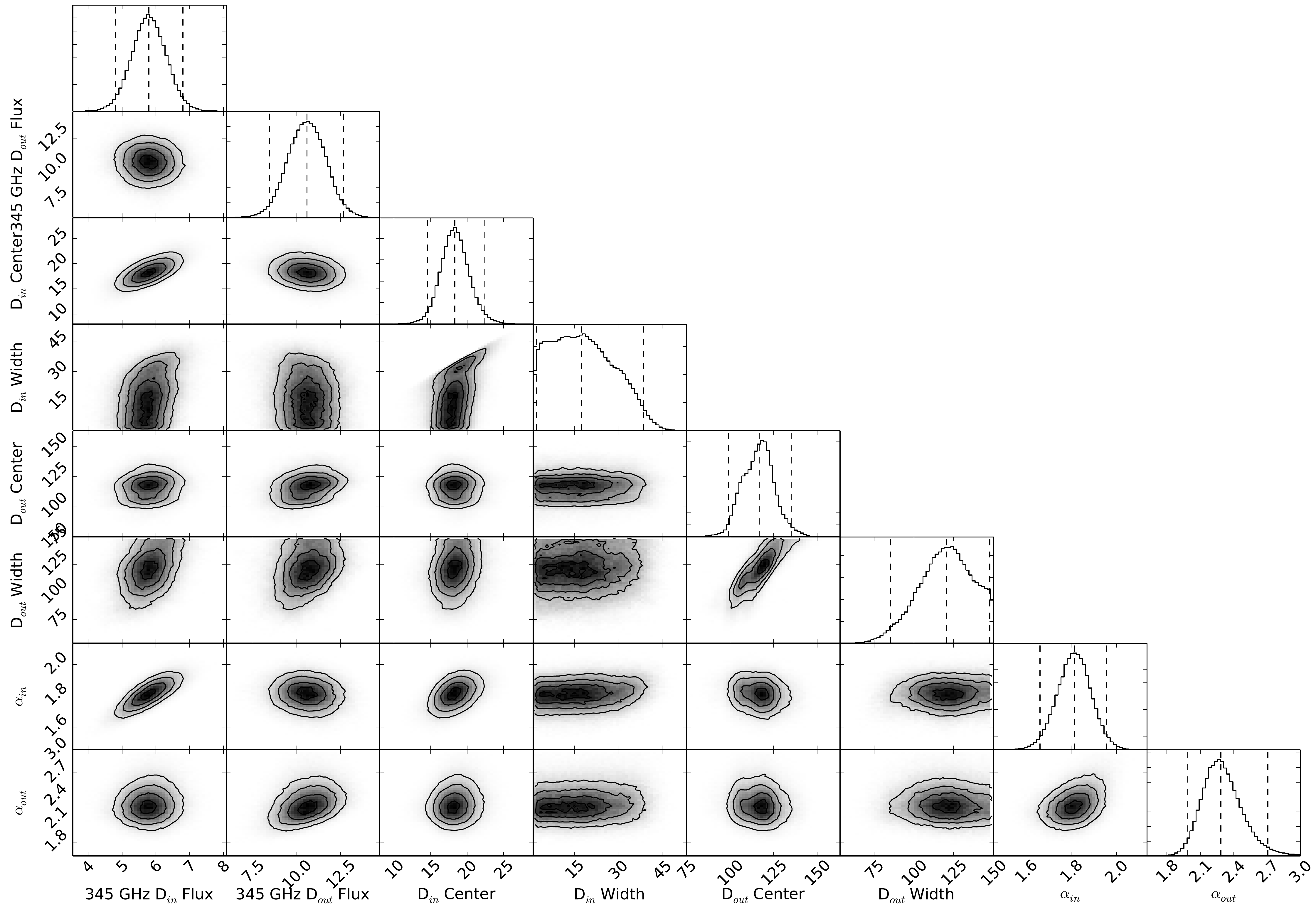}
\caption{MCMC posterior distributions for the two-component model. The dashed lines indicate the most probable values and the 95\% Credible Region. 
\label{pdf2}}
\end{figure}

The two-component model assumes that the system can be characterized by two separate disks. The model again assumes a surface brightness profile $\propto r^{-1}$. The model sets the 345 GHz flux, inner disk spectral index, inner disk center, inner disk width, outer disk center, outer disk width, and outer disk spectral index as free parameters. The flux and spectral index priors are the same as for the single-component model. The inner disk prior range is from 0.1 to 100 and the outer disk prior range is from 50 to 500 au, and models are not accepted if the two disks overlap. The best fit model is given by the magenta curve in Fig.\,\ref{vis1} and the posteriors are shown in Fig.\,\ref{pdf2}. A summary of the most probable model parameters, 95\% Credible Region, and reduced $\chi^{2}$ at each frequency are given in Table\,\ref{obs}.

The resulting inner disk extends from 16 to 45 au with a spectral index of $1.81$. The disk morphology is consistent with both SED models, some scattered light observations \citep{currie}, and ALMA observations \citep{white16}. The spectral index is also consistent with the single-component model and previous estimates \citep{white17}. The outer disk extends from 95 to 300 au with a spectral index of $2.28^{+0.43}_{-0.29}$. This component was not detected in previous analyses of ALMA or VLA data \citep{macgregor, white17}, but is consistent with the location of the spiral features seen in multiple scattered light observations. The spectral index of the outer disk is notably steeper than that of the inner disk. For the inner disk, the best fit 345 GHz flux is 5.8 mJy and the spectral index gives fluxes of 2.8 and 0.62 mJy at 230 and 100 GHz, respectively. For the outer disk, the best fit 345 GHz flux is 11 mJy and the spectral index gives fluxes of 4.2 and 0.63 mJy at 230 and 100 GHz, respectively (the uncertainties as characterized by a 95\% Credible Region are given in Table\,\ref{obs}).

\begin{table*}
\caption{Summary of best fit parameters for the single-component and two-component models. The most probable value for each parameter along with the 95\% credible region from the posterior distribution are given. The reduced $\chi^{2}$ for each model at each frequency is given, along with the average for each model (assuming equal weighting). }
\centering 
\begin{tabular}{c || c | c || c | c } 

 & \multicolumn{2}{c||}{\textbf{Single-Component Model} } &\multicolumn{2}{c}{\textbf{Two-Component Model}}\\
\hline\hline 

   	Parameter & Most Probable & 95\% Credible Range & Most Probable & 95\% Credible Range  \\
   	\hline
   	345 GHz D$_{in}$ Flux [mJy] & 4.5 & [3.8, 5.3] & 5.8  & [4.8, 6.8]\\
   	345 GHz D$_{out}$ Flux [mJy]& -    & -            & 11 & [8.4, 13]\\
   	D$_{in}$ Center [au]        & 28 & [24, 34] & 31  & [25, 38] \\
   	D$_{in}$ Width [au]         & 50 & [39, 65] & 30  & [3.7, 65]\\
   	D$_{out}$ Center [au]       & -    & -            & 195 & [170, 230]\\
   	D$_{out}$ Width [au]        & -    & -            & 200 & [140, 250]\\
   	$\alpha_{in}$               & 1.87 & [1.75, 1.99] & 1.81  & [1.66, 1.96]\\
   	$\alpha_{out}$              & -    & -            & 2.28  & [1.99, 2.71]\\
   	\hline\hline
   	345 GHz $\chi^{2}_{red}$& \multicolumn{2}{c||}{2.78} & \multicolumn{2}{c}{1.34}\\
   	230 GHz $\chi^{2}_{red}$& \multicolumn{2}{c||}{0.77} & \multicolumn{2}{c}{0.80}\\
   	100 GHz $\chi^{2}_{red}$& \multicolumn{2}{c||}{1.60} & \multicolumn{2}{c}{1.10}\\
   	\hline
   	Average $\chi^{2}_{red}$& \multicolumn{2}{c||}{1.72} & \multicolumn{2}{c}{1.08}\\
   	\multicolumn{5}{c}{}

\label{obs}
\end{tabular}
\end{table*}

\section{Discussion}

\begin{figure}
\centering
\includegraphics[width=\textwidth]{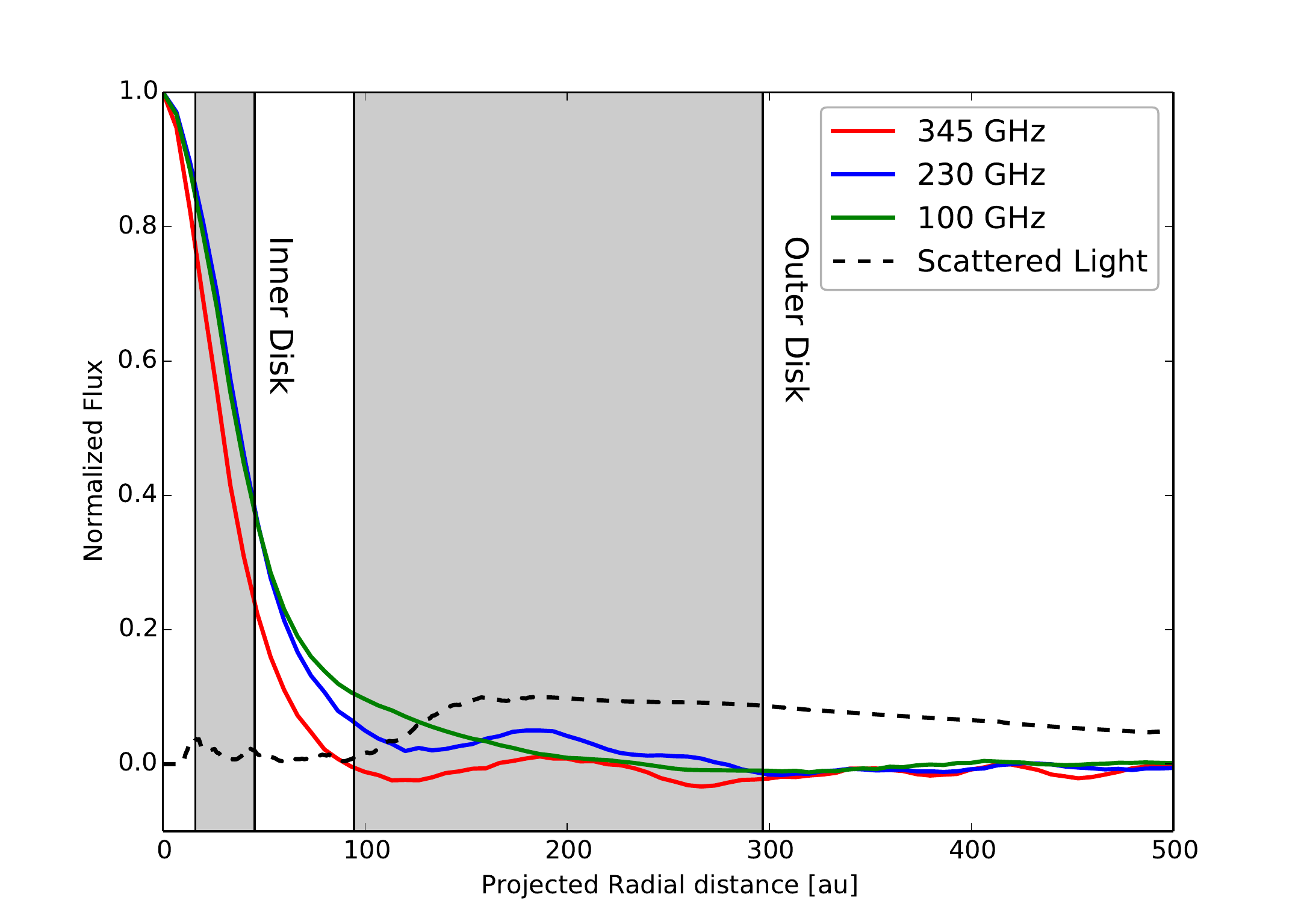}
\caption{Azimuthally averaged radial profiles of HD 141569. The curves were created by summing the total flux in elliptical apertures of inclination 53.4$^{\circ}$ and PA 357$^{\circ}$ \citep{white16}. The red, green, and blue curves are from the ALMA data. They were created from the dirty images at each frequency. The black dashed curve is from the HST scattered light images from \cite{konishi}. The inner disk portion was masked due to artifacts as a byproduct of the PSF subtraction. The most probable locations of the inner and outer disks are marked by the gray shaded areas.
\label{rad}}
\end{figure}

These observations provide the first morphological constraints on mm-sized material outside the inner disk in HD 141569. When comparing the reduced $\chi^{2}$ values for each of the models, the two-component model is preferred (1.74 versus 1.08 averaged reduced $\chi^{2}$). The low reduced $\chi^{2}$ values for both of the 230 GHz models are due to the very large uncertainties on the data points (which are in turn due to the short on source integration time). While the 230 GHz data alone do not strongly prefer one model over the other, all the frequency data together show that a two-component model is at least slightly preferred. 

The inner disk morphology in the two-component model is consistent with SED, scattered light, and ALMA inferred disk properties.  The most probable 345 and 100 GHz fluxes of the inner disk are larger than previously reported at these frequencies \citep{white17}. Prior analyses only fit a single disk to the visibilities with a uniform brightness profile and no central clearing. However, when looking at the visibilities in Fig.\,\ref{vis1}, it is clear that a single-component model can under predict the flux. The spectral index fit to the inner disk is still consistent with the previous estimates. 

The outer disk morphology is broadly consistent with the location of the spiral features seen in scattered light observations. Figure\,\ref{rad} shows azimuthally averaged radial profiles for each ALMA data set image and the HST scattered light images from \cite{konishi} (note that for the scattered light image, the region of the inner disk is unreliable in this plot due to the PSF subtraction creating many image artifacts). These curves were generated by summing up the total flux in elliptical apertures with the inclination and position angle previously constrained in the disk \citep{white16}. The data are then normalized and plotted together in Fig.\,\ref{rad}. The image plane alone does not show a convincing detection of the outer disk material at the location of the spiral arms.  Nonetheless, the slight peak in flux is broadly consistent with the 95 - 300 au outer disk location detected in the visibility data. 

The evidence for the detection of mm emission in the outer disk is clear when you consider that 1) the reduced $\chi^{2}$ is lower for the two-component model; 2) the morphology of the inner disk is only consistent with the shorter wavelength data in the two-component model; and 3) azimuthally averaged radial profiles of the mm data show very tentative evidence of an outer debris disk at a location consistent with both the scattered light observations and the visibility model fitting.

The best fit spectral indicies of $1.81\pm0.15$ and $2.28^{+0.43}_{-0.29}$ for the inner and outer disks, respectively, overlap with expected ranges for both protoplanetary and debris disks. As a protoplanetary disk will likely be optically thick, the spectral index in the Rayleigh Jeans limit will be $\alpha\sim2$. In practice, values of $\alpha$ range from $\alpha = 2.0 \pm0.5$ \citep{andrews} to 1.5-3.2 \cite{ricci11}, but are typically less than 3 \citep{natta}. The spectral indicies observed in debris disks  range from $\sim2-3$ \citep[e.g.,][]{macgregor}. In order to determine whether or not HD 141569 is consistent with a debris origin, properties of the disk other than the spectral index must be considered. It is possible that the grains in HD 141569 represent a mixture of evolutionary stages with elements of both debris and protoplanetary disks.

To approximate the grain size distribution from the spectral index, we adopt the methods of \citet{dalessio}, \citet{ricci}, \citet{macgregor}, and \citet{white_fom}, in which the slope of the size distribution is given by
\begin{equation}
q = \frac{\alpha_{mm} - \alpha_{pl}}{\beta_{s}} + 3,
\end{equation}
where $\beta_{s} = 1.8\pm0.2$ is a power law index for the dust opacity \citep{draine} and $\alpha_{pl}$ is a power law index for the Planck function that depends on the temperature of the dust and the wavelengths of interest \citep[e.g,][]{holland03}.  Specifically,
\begin{equation}
\alpha_{pl} = \left| \frac{\rm log\Big(\frac{B_{\nu_{1}}}{B_{\nu_{2}}}\Big)}{\rm log\Big(\frac{\nu_{1}}{\nu_{2}}\Big)}\right|,
\end{equation}
where B$_{\nu}$ is the Planck Function and $\nu$ is the frequency. 

The inner disk spectral index of $1.81\pm0.15$ corresponds to a grain size distribution of $q =2.95\pm0.10$ (with uncertainties propagated from the 95\% credible region on the spectral index). The outer disk has a spectral index of $2.28^{+0.43}_{-0.29}$ and a corresponding grain size distribution of $q=3.21^{+0.30}_{-0.16}$. The outer disk has a steeper grain size distribution than the inner disk and is at least marginally consistent with some models of debris production \citep[e.g.,][]{pan}. As a reference, a \cite{dohnanyi} collisional cascade will have $q\approx 3.5$.

The presence of a non-negligible amount of gas in the disk, along with the detection of mm grains at large radii, indicates that HD 141569's disk material could be at least partially replenished by a dynamically evolving system of asteroids and comets \citep{matthews}. This is also thought to be the case for other gas-rich debris disks around A stars \citep[e.g., $\beta$ Pic,][]{dent}.  A strict debris interpretation of HD 141569's disk would suggest that, at an age of $\sim5$ Myr, significant grain growth has occurred, producing a large system of planetesimals that are collisionally evolving and repopulating the small grains in the disk. If this view is correct, then systems like HD 141569 and HD 36546  \citep[estimated age range of $3 - 10$ Myr,][]{currie17} show that significant debris generation occurs early in the lifetime of planet-forming disks.

A simple mass estimate for the debris in each component can be calculated by assuming the emission is optically thin, is dominated by density $\rho = 1.0~\rm g~cm^{-3}$ grains, only a single grain size contribute to the emission, the grains are perfect radiators (i.e. albedo $\sim0$), and the disk is populated from the inner to outer edge of each disk. The adopted stellar parameters are $T_{*} = 10500$ K, $L_{*} = 24.2~\rm L_{\odot}$, and $d=111$ pc. The total mass of grains of size s is given by M$_{s} = \frac{4}{3} \pi s^{3} \rho N_{s}$, with N$_{s}$ being the total number of grains of a given size. The number of grains within an annulus at distance r is

\begin{equation}
\frac{dN_{s}}{dr} = \frac{dF_{s}/dr}{B_{\nu}(T_{G}) \Omega_{s}}
\end{equation}

\noindent where $\Omega_{s}$ is the solid angle of a single grain, $B_{\nu}$ is the black body intensity of a grain at temperature $T_{G}= T_{*} \sqrt{\frac{R_{*}}{2r}}$. The disk flux at a single frequency is

\begin{equation}
F_{s} = \int_{R_{in}}^{R_{out}}  2 \pi \sigma_{0} \frac{r_{0}}{r} \frac{r dr}{d^{2}}
\end{equation}

\noindent where $r_{0}$ is a characteristic radius for the surface brightness profile (equal to the inner disk edge in this case), $\sigma_{0}$ is the surface density profile, and $R_{in}$ and $R_{out}$ are the inner and outer edges of the disk. Integrating Eqn.\,3 and solving for the total mass gives

\begin{equation}
M_{s} = \frac{4\sqrt{2}}{9} \frac{s c^{2} d^{2} \rho F_{s}}{\nu^{2} k_{B} T_{*} \sqrt{R_{*}}} \frac{R_{out}^{3/2} - R_{in}^{3/2}}{R_{out} - R_{in}}.
\end{equation}

\noindent For the inner disk, the mass estimates are 0.034, 0.053, and 0.14 $\rm M_{\oplus}$ at 345, 230, and 100 GHz, respectively. For the outer disk the mass estimates are 0.17, 0.22, and 0.37 $\rm M_{\oplus}$. This relation assumes that only a single grain size contributes to a given flux density. The grain radius is set equal to the wavelength of the observations. In practice, multiple grain sizes contribute to emission at a given frequency.

A more comprehensive mass estimate can be made by assuming a full grain size distribution with q for each disk and that the grains only radiate efficiently if their circumference is equal to or larger than the absorbed/emitted photons \citep[see,][]{wyatt, draine}. For this mass calculation, we adopt the methods laid out in \cite{white16, white17} and assume the disk is populated only by grains that have been observed (i.e. $\sim~10~\mu$m to $\sim~3$ mm). Integrating from the inner to outer edge of each disk gives roughly 0.041 $\rm M_{\oplus}$ of small solids in the inner disk and 0.18 $\rm M_{\oplus}$ of small solids in the outer disk. For a strict debris interpretation, the disk should be populated by much larger solids up to the size of asteroids/comets (the limiting size depends on the collision timescales). In reality, the grain size distribution may not follow a single power all the way to these sizes. Extending the current values of q will give unphysically large values for the total mass of the disk (for example we find a M$_{\rm Total}(D<50~\rm km)$ of 75,000 $\rm M_{\oplus}$ and 25,000 $\rm M_{\oplus}$ for the inner and outer disks, respectively).

\section{Summary}

We presented a multi-wavelength analysis of archival ALMA data that shows evidence of mm structure in HD 141569's outer disk. The inner disk is constrained to be within 16 to 45 au and has a spectral index of 1.81. The outer disk is constrained to be within 95 to 300 and has a spectral index of 2.28. The location of the two disk components is consistent with scattered light images. The total disk mass in 10 $\mu$m - 3 mm sized solids is estimated to be 0.041 $\rm M_{\oplus}$ for the inner disk and 0.18 $\rm M_{\oplus}$ for the outer disk. The new constraints on grain properties suggest that HD 141569's outer disk has a steeper grain size distribution than the inner disk.

We thank the anonymous referee for comments that improved this manuscript. J.A.W. and A.C.B. acknowledge support from an NSERC Discovery Grant, the Canadian Foundation for Innovation, The University of British Columbia, and the European Research Council (agreement number 320620). This paper makes use of the following ALMA data: ADS/JAO.ALMA \#2013.1.00883.S, \#2012.1.00698.S and \#2015.1.01600.S. ALMA is a partnership of ESO (representing its member states), NSF (USA) and NINS (Japan), together with NRC (Canada) and NSC and ASIAA (Taiwan) and KASI (Republic of Korea), in cooperation with the Republic of Chile. The Joint ALMA Observatory is operated by ESO, AUI/NRAO and NAOJ.

\newpage

\software{CASA \citep[v4.5.3,][]{casa}}
\facility{ALMA}




\begin{thebibliography}{}

\bibitem[Andrews \& Williams(2005)]{andrews}Andrews, S. M. \& Williams, J. P., 2005, ApJ, 631(2), p.1134

\bibitem[Augereau \& Papaloizou(2004)]{augereau}Augereau, J. C \& Papaloizou, J. C. C. 2004, A\&A, 414, 1153

\bibitem[Brittain \& Rettig(2002)]{brittain} Brittain, S. D. \& Rettig, T. W., 2002 \nat, 418, 57-59 

\bibitem[Clampin et al.(2003)]{clampin}Clampin, M., Krist, J. E., Ardila, D. R., et al., 2003, AJ,
126, 385

\bibitem[Currie et al.(2016)]{currie}Currie, T., Grady, C.A., Cloutier, R., 2016, ApJL, 819(2), p.L26

\bibitem[Currie et al.(2017)]{currie17}Currie, T., Guyon, O., Tamura, M., et al., 2017, ApJL, 836(1), p.L15

\bibitem[D'Alessio et al.(2001)]{dalessio}D'Alessio, P., Calvet, N., Hartmann, L. 2001, ApJ, 553, 321

\bibitem[Dent et al.(2014)]{dent}Dent, W. R. F., Wyatt, M. C., Roberge, A., et al. 2014, Science, 343, 1490

\bibitem[Dohnanyi(1969)]{dohnanyi}Dohnanyi, B. T., 1969, J. Geophys. Res., 74, 2531

\bibitem[Draine(2006)]{draine}Draine, B. T., 2006, ApJ, 636, 1114

\bibitem[Konishi et al.(2016)]{konishi}Konishi, M., Grady, C. A., Schneider, G., et al. 2016, ApJL, 818, L23


\bibitem[Flaherty et al.(2016)]{flaherty}Flaherty, K. M., Hughes, A. M., Andrews, S. M., et al., 2016, ApJ, 818, 97

\bibitem[Goto et al.(2006)]{goto}Goto, M., Usuda, T., Dullemond, C. P., et al. 2006, \apj, 652, 758

\bibitem[Holland et al.(2003)]{holland03}Holland, W. S., Greaves, J. S., Dent, W. R. F., et al., 2003, \apj, 582(2), p.1141

\bibitem[Van Den Ancker et al.(1998)]{ancker}Van Den Ancker, M. E., De Winter, D., Tjin a Djie, H. R. E., et al. 1998, A\&A, 330, 145

\bibitem[Maaskant et al.(2015)]{maaskant}Maaskant, K. M., de Vries, B. L., Min, M., et. al., 2015, A\&A, 574, A140

\bibitem[MacGregor et al.(2016)]{macgregor}MacGregor, M. A., Wilner, D. J., Chandler, C., et al., 2016, ApJ, 823(2), p.79

\bibitem[Matthews et al.(2014)]{matthews}Matthews, B. C., Krivov, A. V., Wyatt, M. C., et al., 2014. Protostars and Planets VI, 521

\bibitem[McMullin et al.(2007)]{casa}McMullin, J. P., Waters, B., Schiebel, D., et al., 2007, Astronomical Data Analysis Software and Systems XVI (ASP Conf. Ser.376), ed. R. A. Shaw, F. Hill, \& D. J. Bell (San Francisco, CA: ASP), 127

\bibitem[Natta et al.(2007)]{natta}Natta, A., Testi, L., Calvet, N., 2007, in Protostars and Planets V, ed. B. Reipurth, D. Jewitt, \& K. Keil (Tucson, AZ: Univ. of Arizona Press), 767

\bibitem[Nillson et al.(2010)]{nilsson}Nilsson, R., Liseau, R., Brandeker, A., et al., 2010, A\&A, 518, A40

\bibitem[Pan \& Sari(2005)]{pan}Pan, M. \& Sari, R. 2005, Icarus, 173, 342

\bibitem[P\'ericaud et al.(2017)]{pericaud}P\'ericaud, J., Di Folco, E., Dutrey, A., et al., 2017, A\&A, 600, p.A62

\bibitem[Ricci et al.(2011)]{ricci11}Ricci, L., Mann, R. K., Testi, L., et al., 2011, A\&A, 525, p.A81

\bibitem[Ricci et al.(2012)]{ricci}Ricci, L., Testi, L., Maddison, S. T., et al., 2012, A\&A, 539, p.L6

\bibitem[Richert et al.(2018)]{richert}Richert, A. J., Lyra, W., \& Kuchner, M. J., 2018, \apj, 856(1), p.41

\bibitem[Thi et al.(2014)]{thi}Thi, W. F., Pinte, C., Pantin, E., et al., 2014, \aap , 561, A50

\bibitem[Weinberger et al.(2000)]{weinberger}Weinberger, A. J., Rich, R. M., Becklin, E. E., et al., 2000, ApJ, 544, 937

\bibitem[White et al.(2016)]{white16} White, J. A., Boley, A. C., Hughes, A. M., et al., 2016, \apj, 829(6), p.11

\bibitem[White et al.(2017a)]{white_fom} White, J. A., Boley, A. C., Dent, W. R. F., et al., 2017, MNRAS, 466(4), pp.4201-4210

\bibitem[White et al.(2017b)]{white17} White, J. A., Boley, A. C., MacGregor, M. A., et al., 2017, MNRAS, 474(4), p.4500 

\bibitem[Wyatt \& Dent(2002)]{wyatt} Wyatt, M. C., \& Dent, W. R. F., 2002, MNRAS, 334, 589

\bibitem[Zuckerman et al.(1995)]{zuckerman95}Zuckerman, B., Forveille, T., \& Kastner, J. H., 1995, Nature, 494-496

\end{thebibliography}
\end{document}